\documentclass[aip,jcp,reprint]{revtex4-1}

\usepackage{dcolumn}
\usepackage{bm}
\usepackage{graphicx}
\usepackage{amsmath, amsthm, amssymb, amsbsy}
\usepackage{color}
\usepackage{ulem}
\usepackage{hyperref}

\begin{document}

\title{Molecular Dynamics Simulation Study of Nonconcatenated Ring Polymers in a Melt: II. Dynamics}

\author{Jonathan D. Halverson}
\affiliation{Max Planck Institute for Polymer Research, Ackermannweg 10, 55128 Mainz, Germany}

\author{Won Bo Lee}
\affiliation{Max Planck Institute for Polymer Research, Ackermannweg 10, 55128 Mainz, Germany}
\affiliation{Department of Chemical and Biomolecular Engineering, Sogang University, Shinsu-dong 1, Mapo-gu, Seoul, Korea}

\author{Gary S. Grest}
\affiliation{Sandia National Laboratories, Albuquerque, NM 87185, USA}

\author{Alexander Y. Grosberg}
\affiliation{Department of Physics, New York University, 4 Washington Place, New York, NY 10003, USA}

\author{Kurt Kremer}
\altaffiliation{The following article has been accepted by \textit{The Journal of Chemical Physics}. After it is published, it will be found at \url{http://jcp.aip.org}. Corresponding author e-mail: kremer@mpip-mainz.mpg.de}
\affiliation{Max Planck Institute for Polymer Research, Ackermannweg 10, 55128 Mainz, Germany}

\date{\today}

\begin{abstract}
Molecular dynamics simulations were conducted to investigate the
dynamic properties of melts of nonconcatenated ring polymers and compared to melts of linear polymers. The longest rings were composed of $N=1600$ monomers per chain which corresponds to roughly 57 entanglement lengths for comparable linear polymers. 
The ring melts were found to diffuse faster than their linear counterparts, with both architectures approximately obeying a $D \sim N^{-2.4}$ scaling law for large $N$. The mean-square displacement of the center-of-mass of the rings follows a sub-diffusive behavior for times and distances beyond the ring extension $\langle R_g^2 \rangle$, neither compatible with the Rouse nor the reptation model. The rings relax stress much faster than linear polymers and the zero-shear viscosity was found to vary as $\eta_0 \sim N^{1.4\pm0.2}$ which is much weaker than the $N^{3.4}$ behavior of linear chains, not matching any commonly known model for polymer dynamics when compared to the observed mean-square displacements. These findings are discussed in view of the conformational properties of the rings presented in the preceding paper\cite{halverson_part1}. [DOI: 10.1063/1.3587138]
\end{abstract}

\maketitle

\section{Introduction}
\label{sec:1}
Understanding the static and dynamic properties of ring polymer
melts is one of the remaining challenges in polymer science. Unlike linear polymers,
the topological constraints for ring or cyclic polymers are permanent and this affects both their
static and dynamic properties. For linear polymers the topological
constraints imposed by the non-crossability of the chains
(entanglements) force each chain to diffuse along its own
coarse-grained backbone, the so-called primitive path, and this is well
described by the reptation model of de Gennes and Edwards. For
branched systems strands have to fold back in order to find a new
conformation without crossing any other chain, resulting in an
exponential growth of the longest relaxation time due to the entropy
barrier of $\cal O(\mathrm{strand~length})$ between different
states\cite{deGennes71,deGennes79,Doi86}. A number of simulation and
experimental results confirm this
concept\cite{Tobolsky,Onogi70,Berry68,Casale71,Odani72,Kurt90,Binder95,McLeish02,Everaers04}.
For both linear and branched polymer melts, it is the free chain
ends that make the known relaxation
mechanisms possible. However, in the case of rings there are no
free ends. This raises a number of unanswered questions regarding
the motion and stress relaxation of ring polymer melts.

In the companion paper\cite{halverson_part1} we discussed the difficulties of understanding the equilibrium (static) properties of a melt of nonconcatenated ring polymers. We pointed out that this system presents a formidable challenge to the current polymer theories. It is of considerable current experimental interest\cite{Roovers83, Hild83, Hild87, Roovers88, Tead92, Rubinstein_Nature_2008,nam_2009}, and it is believed to have a significant projection on the problem of chromatin structure and dynamics in chromosomes. We refer the reader to the detailed discussion of these issues in the Introduction section of our paper\cite{halverson_part1}; the large bibliography of relevant references is also in that paper. The difficulty of the melt of rings problem is such that even the equilibrium structure of nonconcatenated rings is highly non-trivial. We have computed a large number of equilibrium characteristics of this system in the previous paper\cite{halverson_part1}.

In the present paper, the second in a series of two, we investigate the dynamic
properties of ring polymer melts using molecular dynamics (MD)
simulation. To our knowledge the only previous MD studies of ring
polymer melts are by Hur et al. \cite{Hur06} and Tsolou et al. \cite{Tsolou10} who considered
polyethylene for short chain lengths (i.e., $N\lesssim
3N_{e,\mathrm{linear}}$), and Hur et al. \cite{Yoon2010} for chains up to about $11 N_{e,\mathrm{linear}}$, where $N_{e,\mathrm{linear}}$ is the entanglement chain
length of linear polymers at the same density.
Here we consider rings and linear melts composed of much longer chains. The
number of monomers per ring ranges from 100 to 1600 or $3N_{e,\mathrm{linear}}$ to
$57N_{e,\mathrm{linear}}$. Measuring chain lengths in terms
of $N_{e,\mathrm{linear}}$ provides a basis to compare rather different
simulational as well as experimental studies. The linear chain melts were simulated using the same model as the rings with chain
lengths varying from 100 to 800 monomers per chain. Such highly entangled melts have very long
relaxation times and a highly parallelized simulation code had to be used.
For the longest rings as many as 2048 IBM Blue Gene/P cores were used for a single simulation.

All details about the model and the setup of the systems can be found in the preceding
paper, where the very same simulations have been analyzed in order to understand the
static properties of nonconcatenated ring polymers. All properties
reported here are exclusively taken from the fully equilibrated part of the data.
The simulation model is briefly presented in the next section. This is followed by a section
on the dynamics as well as a primitive path
analysis \cite{Sathish05} of the ring systems. A
comprehensive discussion of the results is given in Section \ref{discuss} with the key findings of
the work and future challenges presented in the Conclusions section.

\section{Simulation Methodology}
\label{sec:2}
Based on the Kremer-Grest (KG) model \cite{Kurt90}, all beads
interact via a shifted Lennard-Jones potential (WCA or Weeks-Chandler-Anderson potential)
with a cutoff of $r_c=2^{1/6}~\sigma$.
Nearest neighbor beads along the chain in addition interact via the finitely extensible nonlinear
elastic (FENE) potential, and
chain stiffness is introduced by an angular
potential.\cite{Everaers04}
The model parameters are the same for all ring and linear polymer simulations reported in this work.

The natural time scale of the WCA potential is
$\tau=\sigma\sqrt{m/\epsilon}$, where $m$ is the mass of a monomer.
For the simulations we use $T=1.0~\epsilon/k_B$ and
$\rho=0.85~\sigma^{-3}$. This model has been frequently used for
simulation studies of polymer melts and solutions, so that one can
refer to ample information throughout the discussion of the present,
new results. Also, the linear chain data presented in this work is all
new. The ring melts were simulated using 200 polymers of length 100 to
1600 monomers per chain. The linear systems were composed of $2500$ chains of length $N=100$,
250 chains of $N=200$, and 400 chains of length $N=400$ and 800. An
important parameter for our discussion is
the entanglement length for linear polymers in a melt. For the
model employed here this parameter was determined\cite{Everaers04} to be

\begin{equation}
N_{e,\mathrm{linear}}=28 \pm 1 \ . \label{eq:N_e_for_our_model}
\end{equation}

For more details of the model, the setup and the notation and software used see the preceding
paper\cite{halverson_part1}.

\section{Dynamics Results}
\subsection{Mean-Square Displacements of Monomers and Chains}
\label{dynamics_section}
The dynamics of ring polymers pose special problems for a
theoretical description.
Based on the non-trivial structural properties found in
the preceding paper we have to anticipate dynamic properties
not observed in other polymer systems.
While for short rings one can expect that
the Rouse model might describe the dynamics and stress relaxation
fairly well, this problem for longer rings is completely open. From
the static properties one sees that for large $N$ the rings exhibit an
$N$-independent intrachain density, $\rho_{\mathrm{self}}$,
resulting in a scaling of the overall extension as a globule, while
at the same time there is still significant interpenetration between
rings. For a dense, unknotted and nonconcatenated globule one could
expect a Rouse-like dynamics for the beads on short time scales and
an overall diffusion coupled to amoeba-like shape fluctuations,
similar to segregated chains in two dimensions \cite{KKISchmidt90,wittmer2010}.
However, the mutual interpenetration of rings results in a
stronger interchain coupling of the motion. Whether there are
entanglement effects as in linear polymers is not clear since
there are no free ends to reptate. Kapnistos et al.
\cite{Rubinstein_Nature_2008} and Milner and Newhall \cite{milner10}
have developed models for stress relaxation based on a lattice
animal picture. This however requires entropically unfavorable
double-stranded conformations, which are only somewhat close to the
conformations corresponding to the lattice animal scaling regime
of Fig. 2 in the previous paper\cite{halverson_part1}. Thus
since the expected behavior is quite unclear, we present here a
rather comprehensive first analysis of single and collective ring polymer
dynamic properties. The motion of the rings and linear systems is
characterized by the various mean-square displacements (MSD), and for the rings
these results do not provide a consistent picture
in terms of the known theoretical models.

The first quantities to look at are the different mean-square
displacements of monomers and whole chains as a function of time.
The MSD averaged over all the monomers of a chain is $g_1(t)$, the MSD
of the monomers with respect to the center-of-mass of the chain is
$g_2(t)$, and the MSD of the center-of-mass, $\boldsymbol{r}_{CM}$, of the chains is
$g_3(t)$:

\begin{align}
g_1(t) &= \left\langle \left|\boldsymbol{r}_i (t) - \boldsymbol{r}_i (0)\right|^2\right\rangle,\\
g_2(t) &= \left\langle\left|\boldsymbol{r}_i(t) - \boldsymbol{r}_{CM}(t) - \boldsymbol{r}_i(0) + \boldsymbol{r}_{CM}(0)\right|^2\right\rangle,\\
g_3(t) &= \left\langle \left|\boldsymbol{r}_{CM} (t) - \boldsymbol{r}_{CM} (0)\right|^2\right\rangle.
\end{align}

\noindent
In Eqs. (2) and (3), $\boldsymbol{r}_i$ is the position of monomer number $i$ in space.

\begin{figure*}[]
\includegraphics[scale=1]{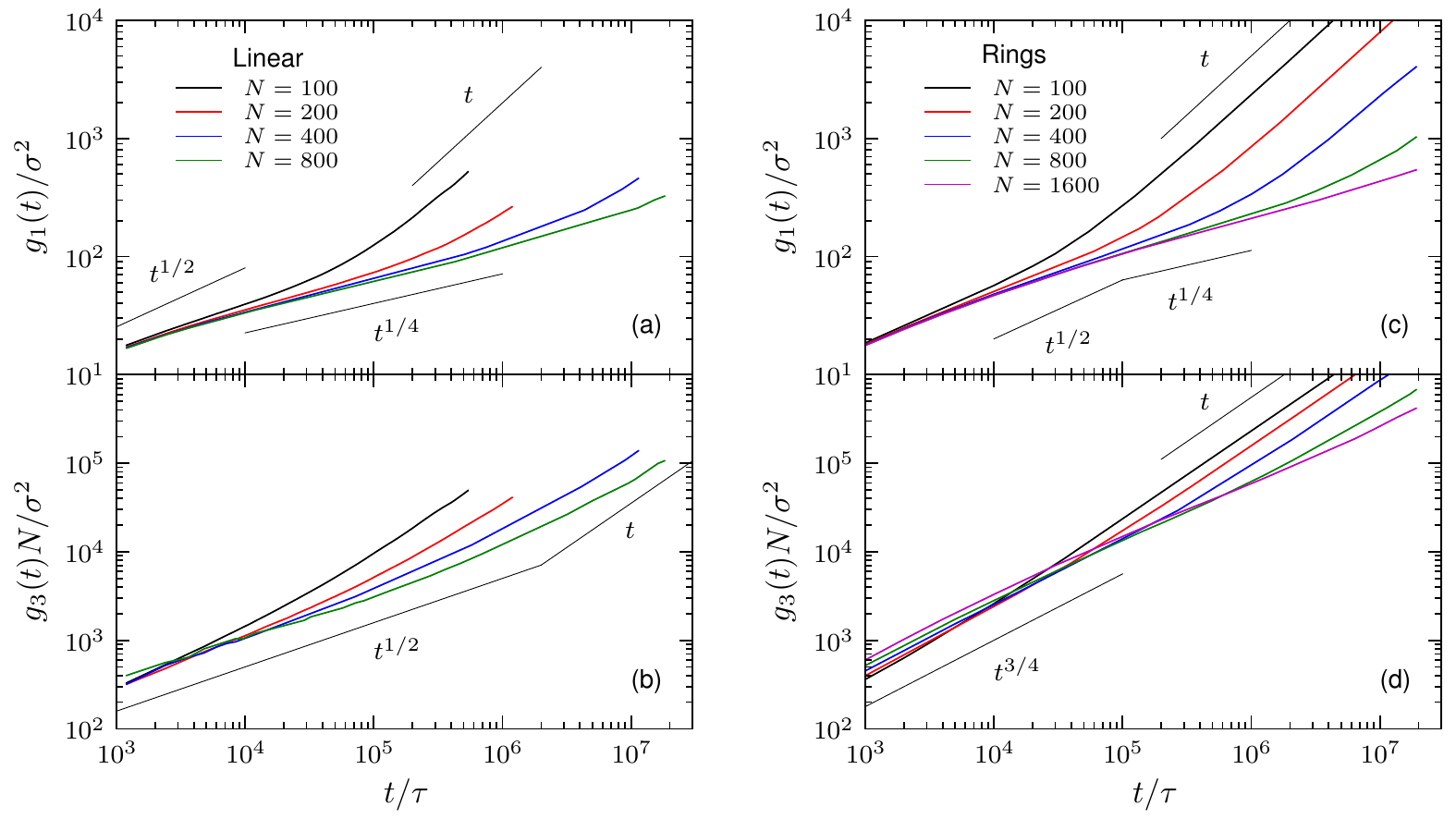}
\caption{(a) $g_1(t)$ and (b) $g_3(t)N$ versus time for the linear systems. These quantities are shown for the rings in (c) and (d). The expected scaling behavior from Rouse and reptation theory is shown along with the $t^{3/4}$ regime for the rings.}
\label{g13_rings_linear}
\end{figure*}

For linear polymers $g_1(t)$ and $g_2(t)$ are averaged over the middle
five beads to reduce chain end effects\cite{Kurt90}. In all cases
the diffusion of the center-of-mass of the whole simulation system
is subtracted. Fig. \ref{g13_rings_linear} shows $g_1(t)$ and $g_3(t)N$ for the ring and
linear systems. Rouse and reptation
theory provide a reasonably good explanation for the behavior of the
linear systems \cite{Doi86,Kurt90}. For the mean-square displacement
of the monomers, $g_1(t)$, the
reptation concept predicts in its asymptotic form \cite{Doi86} a sequence of 
power laws of $g_1 (t) \sim t^{1/2},t^{1/4},t^{1/2},t^1$. For short times the standard Rouse 
behavior dominates. After reaching the Rouse relaxation time $\tau_e$ of an entanglement length $N_e$
the constraints imposed by the surrounding chains cause the beads to follow a Rouse
motion along the random walk contour of the reptation tube, resulting in a $t^{1/4}$ power law. After 
the Rouse relaxation time $\tau_R \sim N^2$  the whole chain diffuses along the tube contour, 
resulting in the second $t^{1/2}$ regime followed by the free diffusion in space after the 
chain leaves the tube, in the ideal model at time $\tau_d \sim N^3/N_{e,\mathrm{linear}}$. This intermediate slowing down
of $g_1(t) \sim t^{1/4}$ comes along with a sub-diffusive regime $g_3(t) \sim t^{1/2}$ for the displacement
of the chain center-of-mass. It should however be noted that the expected  $g_3 (t) \sim t$ in
the early Rouse regime typically reduces to a $t^{0.8}$ power law due to correlation hole effects.

\begin{figure}[]
\includegraphics[scale=1.0]{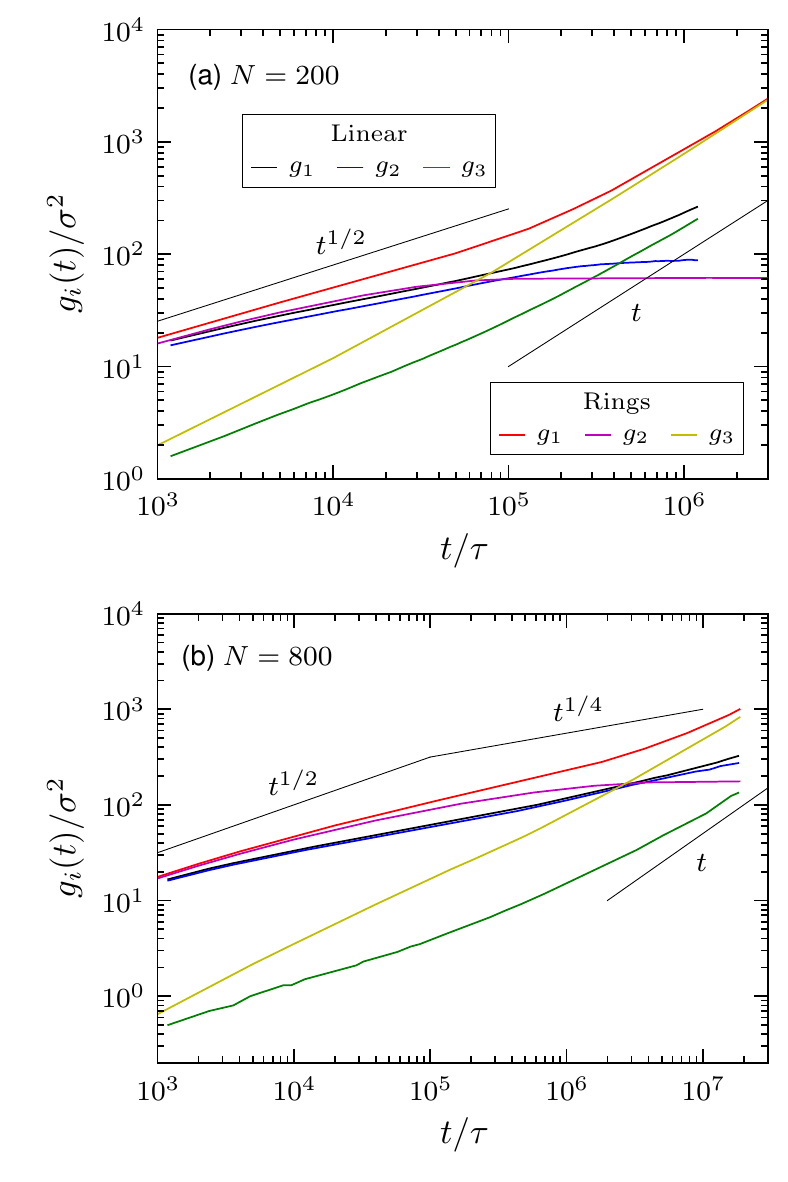}
\caption{The mean-square displacements are shown as a function of time for the rings and linear systems with $N=200$ and 800. The inner monomers are used for the calculation of $g_1(t)$ and $g_2(t)$ for the linear chains. The scaling laws from Rouse and reptation theory are also shown.}
\label{g123_100_800}
\end{figure}

Fig. \ref{g13_rings_linear}a,b gives the results for the linear
polymer melts for chains up to $N=800$. With an estimated
$N_{e,\mathrm{linear}} = 28 \pm 1$ we expect a well developed
reptation-like behavior for the longer chains. We observe a
Rouse-like behavior for the shorter chains and a clear intermediate
regime in $g_1(t)$ with an exponent close to $t^{0.25}$ separating
the two $t^{1/2}$ regimes in $g_1(t)$ for the long chains.  There is
the well known very smooth crossover, making it very difficult to
precisely identify the crossover times for the different regimes. Fig.
\ref{g13_rings_linear}b shows $g_3(t)N$ versus time for the linear
systems. As expected from the Rouse description there is an
immediate crossover towards free diffusion for the shorter chains
whereas the longer ones exhibit a well-defined $t^{1/2}$ regime
before crossing over into free diffusion (with some small deviations for 
short times for $N=800$). In all cases the free
diffusion of the center-of-mass of the chains sets in long before
$g_3(t)$ reaches the radius of gyration squared, i.e., at about
$\left\langle R_g^2 \right\rangle / 2.5$ for $N=200$ and $\left\langle R_g^2
\right\rangle /3.0$ for $N=800$. Though there are still many details
to be understood in the context of the dynamics of linear polymers,
the Rouse and reptation theories give a very good semi-quantitative
description.

While for linear polymer melts the basic dynamic concepts are fairly
clear, the situation for ring melts is strikingly different. Fig.
\ref{g13_rings_linear}c shows $g_1(t)$ for the rings. All five data
sets follow the same curve during early times. Then, independent of
$N$ we observe a significant slowing down in the  mean-square
displacement from the initial  $t^{1/2}$ scaling to a power law very
close to the $t^{1/4}$ expected for perfect reptation of linear
polymers. Note however that the rings cannot entangle in the
classical sense and thus the rings are not expected to show
reptation-like behavior. A scaling regime of $g_1(t) \sim t^{0.35}$ has been reported for melts of polyethylene rings\cite{Yoon2010} of length $11N_{e,\mathrm{linear}}$ which agrees with the present findings.
After the approximate $t^{1/4}$ regime we do not observe a second $t^{1/2}$ regime in $g_1(t)$ at a value
of ${\cal O}(N^{1/2})$ (much smaller than the radius of gyration squared),
but rather a direct crossover toward the diffusion of the whole
chain. From Fig. \ref{g13_rings_linear}c we observe
this crossover at a displacement significantly larger than $\left\langle
R_g^2 \right\rangle^{1/2}$, something \textit{not observed in any other polymeric
system}! The crossover times are around $t\approx 7.6\times10^4~\tau$,
$2.4\times10^5~\tau$, $1.5\times10^6~\tau$, $8.2\times10^6~\tau$ and $g_1(t) \approx
150~\sigma^2$, $200~\sigma^2$, $330~\sigma^2$, $450~\sigma^2$ for
$N=100$, 200, 400 and 800, respectively. For the $N=1600$ 
case, no onset of a crossover
towards diffusion is seen for up to $2.0\times10^7~\tau$.
In Fig. \ref{g13_rings_linear}d, $g_3(t)N$ is shown for the rings
along with a reference line with a slope of 3/4. This scaling
behavior is seen to provide reasonable agreement with the
center-of-mass motion of all five ring systems
and for the $N=400$ system it holds for several decades. The $t^{3/4}$ power law reminds one of a correlation hole
effect for Rouse chains, rather than any reptation phenomenon. The
crossover towards free diffusion is in good agreement to what is
observed from $g_1(t)$. However, the crossover to the free diffusion
time scaling of $t^1$ occurs at values significantly larger than the
chain extensions (cf. Table 2  of the preceding paper), similar to what is seen for $g_1(t)$.  
This is in contrast to
the linear systems which show a $t^1$ scaling after moving
even somewhat less than their own size. For the longest rings the fully diffusive regime is not reached
within the time frame of the simulation.
For $N=1600$ eventually we observe an even smaller effective power law, 
deviating from the $t^{3/4}$ towards a minimal power around $t^{0.65}$. The mean-square displacement
data for the rings and linear systems is available in the supplementary material\footnote{See ancillary files for the mean-square displacement data of the rings and linear systems as well as the various relaxation times of the rings.}.

Another striking observation from Fig. \ref{g13_rings_linear}d is that $g_3(t) N$ for short times does not
become independent of $N$. It displays a small but still clearly visible
ring length dependence, indicating a delayed onset of the constraints
imposed by surrounding rings with increasing ring size.
Such phenomena have not been
observed in a polymeric system so far and suggests long-range
ring-ring correlations, even though the chains cannot be entangled
in the classical sense.
The slight increase of $g_3(t) N$ for short times
and increasing $N$ is indicative of the fact that crumpled globules, which
is the presumed state of nonconcatenated rings in the melt,
are characterized by a rather deep correlation hole\footnote{For very short times,
below the ones shown here, $g_3(t) N$ is independent of $N$ and $\sim t^1$.
Thus the observed behavior indicates that the initial free diffusion extends
to slightly larger distances for the longer rings, reminiscent of typical cage effects
in liquids.}.

To gain additional insight into these correlations we examined the dynamics of the exchange of neighbors. The neighbors of a given chain are defined as the other chains whose center-of-mass is within a distance $\left\langle R_e^2 \right\rangle^{1/2}$ of the center-of-mass of the reference chain. In the preceding paper\cite{halverson_part1}, we denoted the average equilibrium value of this quantity as $K_1(r)$ with $r=\left\langle R_e^2 \right\rangle^{1/2}$ or the root-mean-square spanning distance between beads $N/2$ apart along the ring. At any given time each chain has a specific set of neighbors with the average number of chains in the set being $K_1$, which is 11.6 for $N=100$ and 16.7 for $N=1600$. We computed the number of remaining (unexchanged) neighbors after the reference ring moved its own root-mean-square gyration radius. This number was found to increase with increasing chain length from $5.5$ for $N=100$ to $7.3$ for $N=1600$. We also computed how far the reference chain moved before all neighbors but one were exchanged. The calculation was performed using a large number of time origins and the root-mean-square displacement was found to increase with increasing chain length from about $2.5 \left\langle R_g^2 \right\rangle^{1/2}$
for $N=100$ to significantly more than
3$\left\langle R_g^2 \right\rangle^{1/2}$ for $N=800$. It is interesting to note that these displacements are roughly the same as
those required for the systems to reach the fully diffusive regime.

\begin{figure}[]
\includegraphics[scale=1.0]{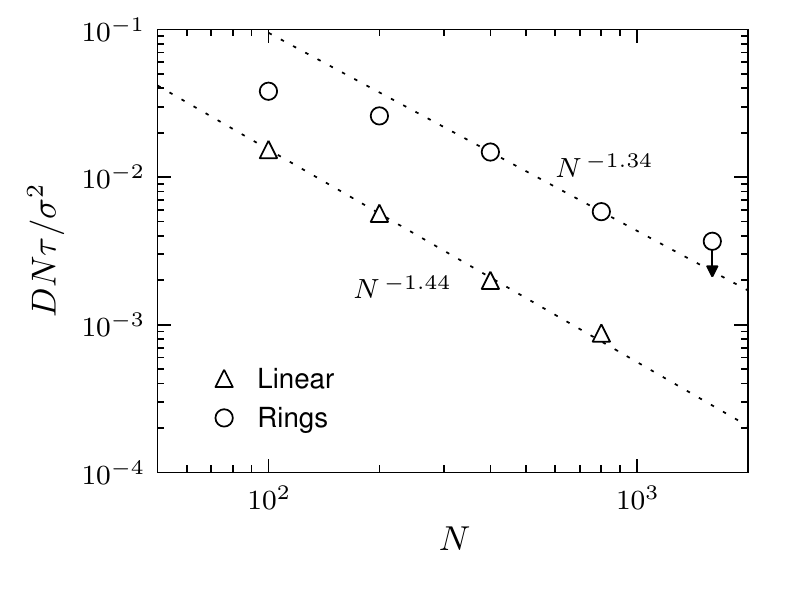}
\caption{Self-diffusion coefficient multiplied by $N$ versus chain length for the rings and linear melts. The line through the ring
data was fit to the values for the $N=400$ and 800 systems while for the linear data all points were used. The arrow beneath the point
for the largest ring system indicates that its value is expected to decrease somewhat
until the system reaches the fully diffusive regime. The range of this shift (length of arrow) is based on the extrapolation
of $g_3(t)/t$ to infinite time. With the exception of the $N=1600$ ring case, all error bars are smaller than the size of the symbols.}
\label{diff_coeff}
\end{figure}

\begin{figure}[]
\includegraphics[scale=1.0]{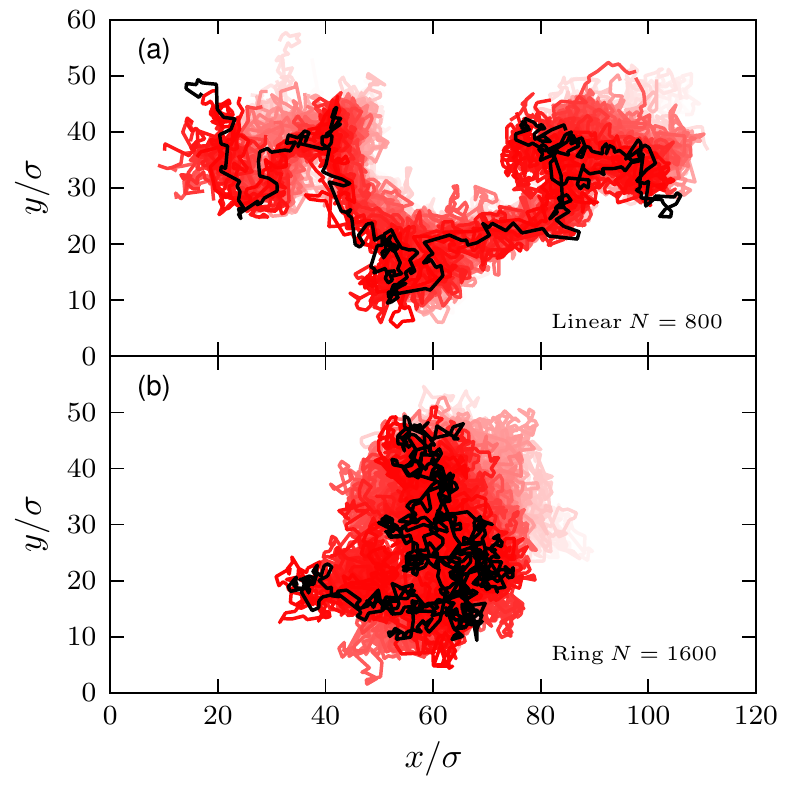}
\caption{(a) Linear chain with $N=800$ shown every $12000~\tau$ for 60 conformations or $0.28\tau_{R,\mathrm{linear}}$ and (b) ring with $N=1600$ shown every $10000~\tau$ for 60 conformations or $0.23\tau_{R,\mathrm{rings}}$. Chain snapshots at later times are drawn darker in color with the final chain conformation in black.}
\label{nature_of_motion}
\end{figure}

Even though the ring polymers display a remarkable and so far unseen
slowing down compared to a Rouse-like motion, they diffuse
significantly faster than their linear counterparts. This is
illustrated in Fig. \ref{g123_100_800}, where the MSD
for two chain lengths are compared
directly for both melts of rings and of linear chains. As expected
$g_2(t)$ and $g_3(t)$ of the rings cross around $2\left\langle R_g^2
\right\rangle$. For the longer rings, however, $g_3(t)$ at this crossing point has not
yet reached the fully diffusive regime.
For the linear polymers the
entanglement effects are much stronger, and the crossing of $g_2(t)$
and $g_3(t)$ is hardly reached. It should however be noted that this
crossover for the linear chains is below $2\left\langle R_g^2 \right\rangle$
and occurs in a regime where $g_3(t)$ already displays
the standard diffusion behavior. The self-diffusion coefficients, as
determined by the long-time behavior of $g_3(t)$,
are shown in Fig. \ref{diff_coeff} for the rings and linear systems.
The rings diffuse faster than the linear systems for all chain
lengths considered. This is in agreement with previous simulation
results \cite{Muller96,Brown98b,Muller00}. And while early
experimental papers \cite{Hild87, Tead92} report that the two
architectures have similar diffusivities, recent experiments cast
doubt on these findings \cite{Rubinstein_Nature_2008}. Reptation
theory predicts for linear polymers $D \sim N^{-2}$ for
$N>N_{e,\mathrm{linear}}$, however, in practice the observed scaling \cite{Lodge1999}
is $N^{-2.3 \pm 0.1}$ and this is found here to within the error bars.
The data for the ring systems
are almost parallel (in log-log scale), but shifted to
significantly faster diffusion. Using the values from the $N=400$
and 800 cases one finds $D \sim N^{-2.3}$ for the rings. This
is a stronger scaling than reported previously for shorter rings. One reason for this
is that the present study uses longer chains with
$N/N_{e,\mathrm{linear}} \approx 57$ for $N=1600$. Despite a
simulation time of more than $3.9\times10^7~\tau$, the longest ring system did
not reach the fully diffusive regime. The arrow in Fig.
\ref{diff_coeff} indicates the extrapolated final value which will be
slightly lower than the data point. Because the linear systems
follow the reptation picture for chains as short as $N=100$, the
effective entanglement length for the rings must be higher than that of the linear chains, if the
data were to be interpreted in terms of some as yet unknown
generalization of the reptation concept.
The corresponding
entanglement length for the rings, $N_{e,\mathrm{rings}}$, will be
estimated in Section \ref{ppa_sec}.
This is all to be compared to a visualization of the chain motion,
which significantly differ between the two architectures.
Fig. \ref{nature_of_motion} compares the
motion of a linear chain to that of a ring for times less than $\tau_R$.
The linear chain is found to be largely confined
to a tube with the end monomers having more freedom than the inner
monomers. This is in sharp contrast to the ring in Fig. \ref{nature_of_motion}b which is found to trace out
a roughly spherical volume.

\begin{figure}[]
\includegraphics[scale=1.0]{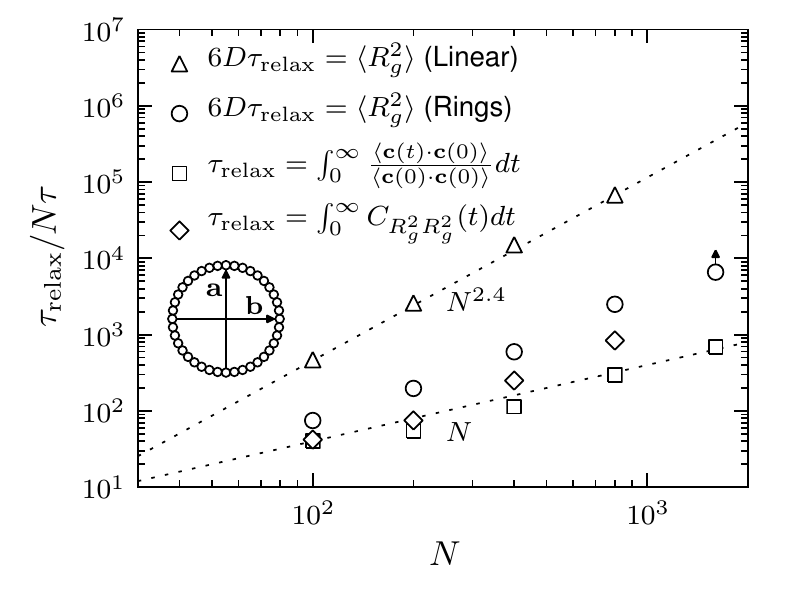}
\caption{The diffusive and conformational relaxation times divided by $N$ as a function of chain length. Inset: The $\boldsymbol{a}$ vector is formed between monomers $i$ and $i + N/2$ while $\boldsymbol{b}$ is formed from $i + N/4$ and $i + 3N/4$. With $\boldsymbol{c}=\boldsymbol{a} \times \boldsymbol{b}$, the time scale for single-chain structural relaxation is found by integrating the normalized autocorrelation function of $\boldsymbol{c}$. Despite a very long simulation time it was not possible to accurately determine the correlation time of $C_{R_g^2R_g^2}(t)$ for the $N=1600$ ring system. The error bars for the relaxation times of the ring systems with $N < 1600$ are smaller than the symbol sizes. The line through the linear chain data is a fit while the $\tau_{\mathrm{relax}}/N \sim N$ line is for reference only.}
\label{relaxation_time}
\end{figure}

The slow crossover towards free diffusion of the rings is quite striking. For
melts of strictly two-dimensional polymers it is known that the chains segregate
and form 2-d globular structures. The diffusion is mostly due to amoeba-like shape
fluctuations, which do not require the complete reorganization of the internal
winding of the chain. In a similar way we can check whether the diffusion
time of a ring, the correlation time of the radius of gyration squared and
the internal contour diffusion follow the same time dependency
or not. If $\boldsymbol{a}$ and $\boldsymbol{b}$ are spanning vectors whose tails are separated by $N/4$
monomers (see inset of Fig. 5) then $\boldsymbol{c} = \boldsymbol{a} \times \boldsymbol{b}$.
In practice, the points forming the head and tails of the spanning vectors are averaged over five adjacent monomers. The correlation time of the
autocorrelation function $C_{\boldsymbol{cc}}(t)$ provides a time scale for the internal structural
rearrangements of the rings. Correlation times are computed by numerically
integrating the normalized $C_{\boldsymbol{cc}}(t)$ from zero to infinity. These times are compared
to the diffusion times and the correlation time of the
radius of gyration squared in Fig. \ref{relaxation_time}. The longer rings are found to
undergo substantial internal rearrangements much faster than they diffuse their
own size or relax the overall extension. Not only is the amplitude different,
but also the exponent in the power law dependence of $N$ on $\tau_{\mathrm{relax}}$ is
significantly lower and approaches a value close to 2.2. Thus the overall
diffusion decouples from the local internal rearrangement, as implied by the decay
of $C_{\boldsymbol{cc}}(t)$. This strongly suggests that the slowing down in the mean-square
displacement of the overall chain, $g_3(t)$, is due to ring-ring coupling, which
then also dominates the overall relaxation of $R_g^2$.
For $N \geq 400$ the correlation times of $R_g^2$ are found to
be significantly larger than the internal reorganization times and they closely
follow the time scaling of the overall diffusion of the chains. This implies that
internal structural rearrangement takes place with a fairly small change in the
overall extension of the chain.

\begin{figure}[]
\includegraphics[scale=1.0]{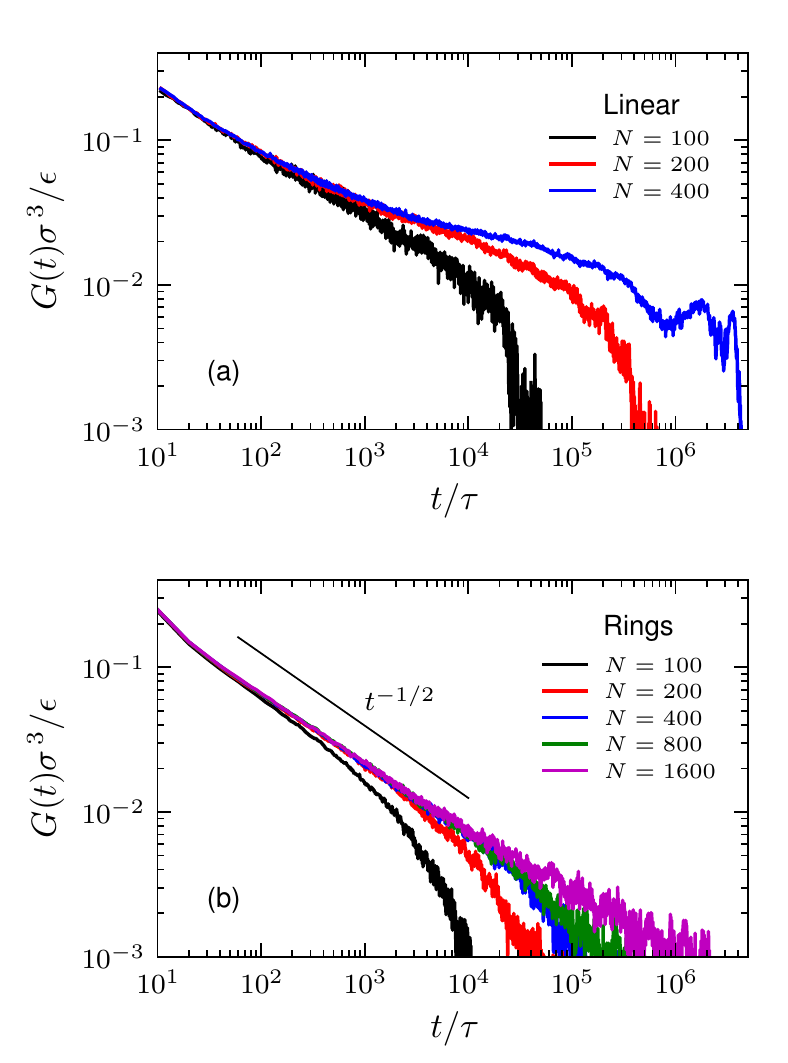}
\caption{Stress relaxation for the (a) linear and (b) ring systems. The Rouse scaling is shown for reference in (b). $G(t)$ was computed as described in the main text but additional averaging was performed to make the curves more discernable. The error bars for $G(t)$ increase with increasing $t$.}
\label{stress_relaxation_ab}
\end{figure}

If the internal reorganization time and the diffusion time are that
different, this might also have consequences for the viscosity of
the melts. The shear relaxation modulus is
\begin{eqnarray}
G_{\alpha \beta}(t)=\frac{V}{k_BT}\big\langle \bar{\sigma}_{\alpha\beta}(t)
\bar{\sigma}_{\alpha \beta}(0)\big\rangle,
\label{stress_eqn}
\end{eqnarray}
\noindent where $\bar{\sigma}_{\alpha \beta}(t)$ is the pre-averaged
stress \cite{Lee09}. In Eq. \ref{stress_eqn}, $\alpha$ and $\beta$ are Cartesian indices with $\alpha \ne
\beta$, and $V$ is the volume of the simulation box. $G(t)$ is the
average of $G_{\alpha \beta}(t)$ with $\alpha \beta = xy,\, xz,\, yz$.
The shear relaxation modulus is shown in Fig. \ref{stress_relaxation_ab}a,b for
the linear and ring systems. The
linear systems show the characteristic behavior. That is, a
Rouse-like decay at early times followed by a plateau region and
finally exponential decay. For the rings the behavior is quite
different. The ring systems show $G(t) \sim t^{-1/2}$ followed by
exponential decay. Specifically, the exponent for the stress
relaxation tends to increase from $-0.5$ to $-0.45$ with increasing $N$. The
functional form and exponent for $G(t)$ are nearly the same as those
suggested by Kapnistos et al.\cite{Rubinstein_Nature_2008} The
ring melts are found to relax stress much more rapidly than the
linear systems. There is no sign of a plateau even for the largest rings,
suggesting that a standard tube model certainly is not applicable.
The difference in noise between the two data sets arises from
differences in averaging schemes and system sizes (cf. Table
1 of the preceding paper). For the linear chains, the stress was computed every
other step for 25 steps and averaged for each 100 step interval. For
the rings the stress was computed at every step and averaged over each
1000 step interval or $10~\tau$.

\begin{figure}[]
\includegraphics[scale=1.0]{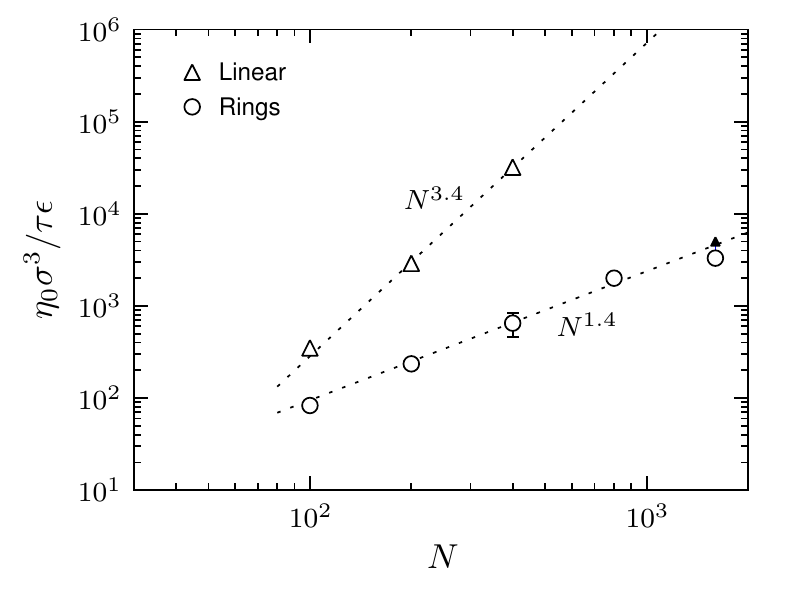}
\caption{Zero-shear viscosity of the rings and linear melts versus $N$. The error bars for the $N=400$ ring case were obtained from four independent simulations while the other points were obtained from single simulations. While it is difficult to estimate the error bars for the $N=1600$ case those of the 800 case are expected to be smaller than twice the symbol size. Note the arrow above the data point for the $N=1600$ system.}
\label{viscosity_figure}
\end{figure}

Results for the zero-shear viscosity obtained from the Green-Kubo
relation \cite{Allen87}, $\eta_0 = \int_0^\infty G(t)dt$, are shown in Fig. \ref{viscosity_figure}. 
The linear systems show a $N^{3.4}$
scaling which is in agreement with previous simulation \cite{Kroeger2000} and
experimental results.\cite{Ferry80} Even though the error bars
become quite large for the longer rings and for $N=1600$ we can only
give a rough estimate of $\eta_0$, the rings show a much weaker dependence on chain length
with $\eta_0 \sim N^{1.4 \pm 0.2}$. For comparison, the Rouse model
predicts a linear $N$-dependence. This is another result which requires 
a deeper understanding. It appears that stress relaxation, which
governs the viscosity, follows a time behavior much
closer to the internal reorganization of the rings rather than
the overall diffusion and relaxation of $R_g^2$. The viscosity measurements
for the rings and linear systems have been confirmed by nonequilibrium
molecular dynamics simulations\cite{tuckerman97}. These results will
appear in an upcoming paper\cite{Halverson10}. The zero-shear viscosities are given for the rings
and linear systems in Table \ref{table3}.

\begin{table}[ht]
\begin{center}
\begin{tabular}{c|cc|cc}\hline& \multicolumn{2}{c|}{Rings} & \multicolumn{2}{c}{Linear} \\
\hline
$N$ & $D \times 10^5$ & $\eta_0$ & $D \times 10^5$ & $\eta_0$ \\\hline
 100  & 38.2     & 83          & 15.4  & 350   \\ 
200  & 13.0     & 235         & 2.85  & 2900  \\ 
400  & 3.7      & $650\pm180$ & 0.50  & 32100 \\ 
800  & 0.73     & $\approx 2000$        & 0.11  &  --   \\
1600  & $<0.23$  & $>3300$     & --    &  --   \\
\hline
\end{tabular}
\end{center}
\caption{Transport coefficients for the rings and linear systems. For the same value of $N$ the rings are found to have a higher diffusivity and a lower viscosity. $D$ and $\eta_0$ are given in units of $\sigma^2/\tau$ and $\tau\sigma^3/\epsilon$, respectively. The diffusion coefficient and the viscosity for the $N=1600$ ring case are upper and lower bounds, respectively, since both have not settled down completely within our analysis window. This is partially due to the extensive averaging needed. The error bars for $\eta_0$ for the $N=400$ ring case were obtained from four independent simulations.}
\label{table3}
\end{table}

As an additional check, a simulation was performed on the $N=800$
ring system using a dissipative particle dynamics\cite{hoogerbrugge,
espanol} (DPD) thermostat\cite{Soddemann03} instead of the Langevin
thermostat. The same friction coefficient was used and $r_c$ was
taken as $2^{1/6}~\sigma$. The simulation was run for $8\times10^5~\tau$ and
it showed the same behavior for the mean-square displacements
indicating that hydrodynamics or momentum conservation are
unimportant for these systems on this time scale.

\subsection{Primitive Path Analysis}
\label{ppa_sec}

Even though ring polymers in a melt cannot entangle in the classical
sense, the slowing down of the overall chain diffusion as well as
the properties of $g_1(t)$ and the slow exchange of neighbors
suggest that neighboring rings must be somehow coupled in order to
prevent fast free diffusion. Thus an adapted primitive path analysis
(PPA) was conducted to elucidate the nature and extent of possible
quasi-entanglements in the ring systems. The procedure for linear
chains \cite{Everaers04} was modified for this purpose. Starting
with an equilibrated configuration, a pair of opposite monomers ($i$
and $i + N/2$) of each ring polymer are chosen at random to be fixed
in space throughout the PPA. With the angular and nonbonded
intrachain interactions switched off, the temperature was lowered to
$0.05~\epsilon/k_B$ and the dynamics were carried out until
equilibrium was reached. Without fixed monomers the rings collapse
to points. At least three cases were run for each
value of $N$. 

\begin{figure*}[]
\includegraphics[scale=1.0]{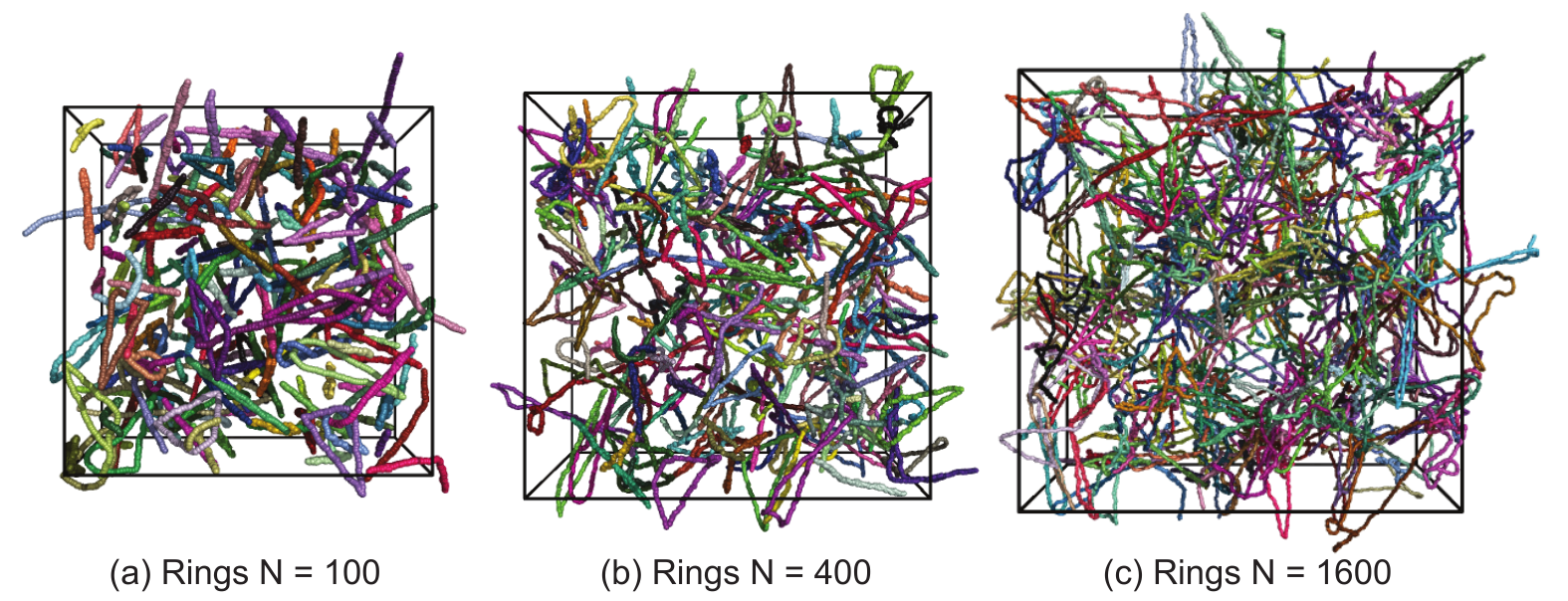}
\caption{Final configurations from a primitive path analysis of the ring systems: (a) $N=100$, (b) 400 and (c) 1600.}
\label{ppa_200_1600}
\end{figure*}

Fig. \ref{ppa_200_1600} shows the last configuration from each PPA
simulation for $N=100$, 400 and 1600. For the $N=100$ case, many of
the chains have collapsed to double-folded lines with an average
contour length, $L_{pp}$, of $2\left\langle R_e \right\rangle$.
Rings that have not completely collapsed are
mutually constrained by one or more neighboring chains. As $N$
increases, the equilibrium fraction of double-folded lines decreases
dramatically. For instance, for the $N=1600$ case shown in Fig.
\ref{ppa_200_1600}c all the chains are mutually quasi-entangled with
not a single double-folded line. It can also be seen for increasing
chain length that there is a greater tendency for rings to thread or
pass through one another. While the chains were initially segregated
at the start of the simulation, at equilibrium they are found to
form a highly connected melt, even though the volume of each ring
is only shared by a small number of other rings.

\begin{figure}[]
\includegraphics[scale=1.0]{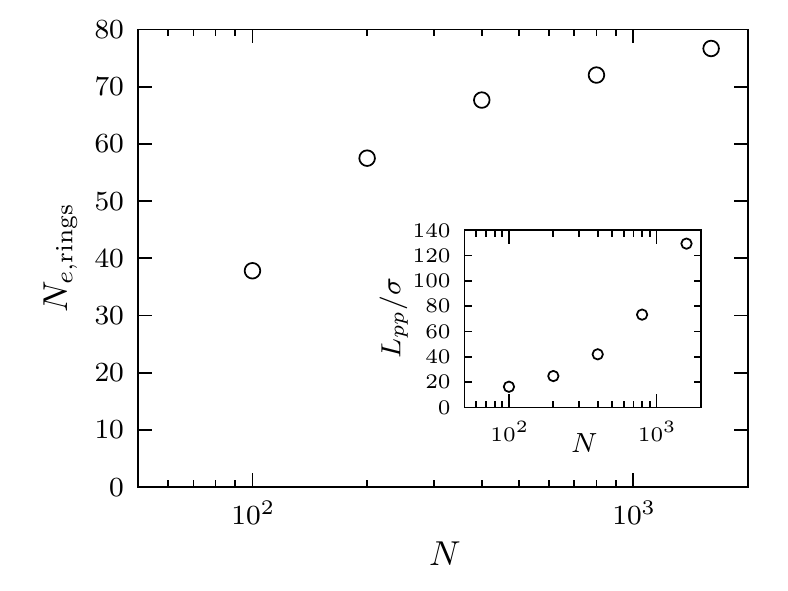}
\caption{$N_{e,\mathrm{rings}}$ as a function of chain length. For our model $N_{e,\mathrm{linear}}=28 \pm 1$. The inset gives the average contour length versus $N$.}
\label{Lpp_n_PPA}
\end{figure}

\begin{figure}[]
\includegraphics[scale=1.0]{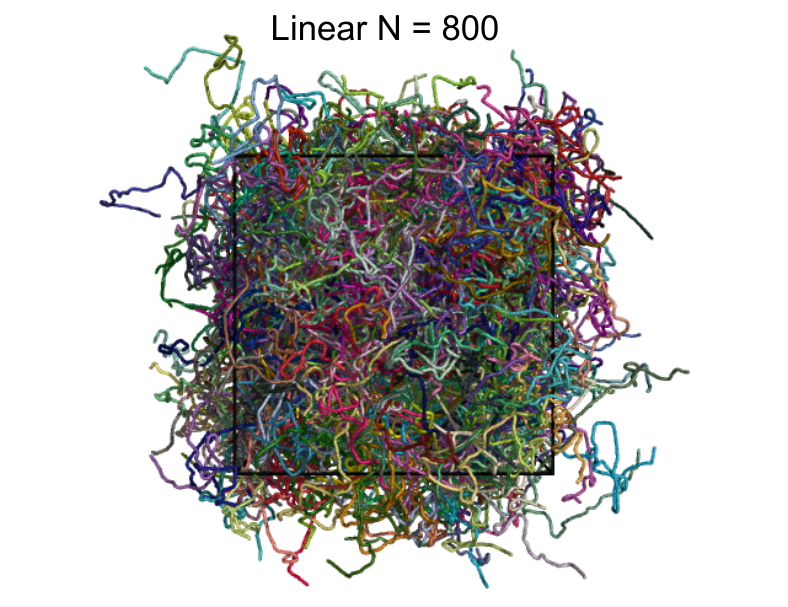}
\caption{Same as Fig. \ref{ppa_200_1600} except for a linear system with $M=400$ and $N=800$. The linear chains with $N=800$ are far more entangled than the $N=1600$ rings (cf. Fig. \ref{ppa_200_1600}).}
\label{PPA_linear}
\end{figure}

While somewhat
questionable here, the standard argument to deduce an entanglement
length for linear chains may be applied to the rings. For the linear
chains the entanglement length is $N\left\langle R_{e,\mathrm{linear}}^2 \right>/L_{pp}^2$, which is 
the number of monomers per Kuhn segment of the primitive path. For the rings this
expression is modified by recognizing that a ring should be treated as two linear chains each of length $N/2$ with common end points.
In Fig.
\ref{Lpp_n_PPA} we plot $N_{e,\mathrm{rings}}$ against chain length
where $N_{e,\mathrm{rings}}=(N/2)\left\langle R_{e,\mathrm{rings}}^2 \right>/(L_{pp}/2)^2$.
For the linear chains one finds $N_{e,\mathrm{linear}} \approx 28$,
which is independent of $N$ for sufficiently long chains. The
entanglement number for the rings is found to vary with chain length
with
the longest rings giving approximately 77.
This value, however, still seems to slowly grow, which is not totally
surprising in view of the early time values of $g_3(t)N$ as shown
in Fig. \ref{g13_rings_linear}d and the slow crossover to an asymptotic value of $K_1$ (see preceding paper\cite{halverson_part1}).
The average contour length for each
ring system is given in the inset to Fig. \ref{Lpp_n_PPA}. The final
configuration from the PPA
for the $N=800$ linear system is shown in Fig.
\ref{PPA_linear} for comparison.
The entanglement density of the
linear system with $N=800$ is higher than that of the $N=1600$ 
ring system. Our computed value of $N_{e,\mathrm{linear}}$ for this system agrees with 
the previously reported value of $28 \pm 1$\cite{Everaers04}.

These results are to be compared with the fact, found in the previous paper, that
the pressure increase due to the entropy loss
associated with the rings squeezing each other in the melt
was too small to be observed in our simulations.  If the concept
of entanglement length $N_{e,\mathrm{rings}}$ makes sense for rings then
it should be possible to estimate this pressure increase as
$1/N_{e,\mathrm{rings}}$ times the reference pressure.  Given the above
estimates of $N_{e,\mathrm{rings}}$ we do not expect to be able to
observe such a small difference in pressures.

\section{Discussion}
\label{discuss}
In the previous paper\cite{halverson_part1} we investigated the structural properties
of a melt of nonconcatenated rings. It was found that the
overall conformations have a number of peculiar properties and are by no
means trivial.
These peculiar structural properties of the long ring polymer melts
have significant consequences for the dynamics. The diffusion of
linear chains is explained by the Rouse and reptation theories as demonstrated in the present work.
These theories make use of the fact that linear chains have free ends but
ring polymers do not have
free ends so the application of these theories, and especially reptation
theory, is questionable. However, the Rouse model can be solved for a Gaussian
ring\cite{Tsolou10}. The diffusion times from
Rouse theory for the linear and ring models are given in Table \ref{table4}.
Even for the smallest value of $N$ only fair agreement
is found between the simulation results and those predicted by
the theory for both architectures. The disagreement grows with increasing $N$ as the
systems become more and more entangled, a phenomena which is completely ignored by the theory.
The Rouse theory prediction for the viscosity of the rings is in better but also not perfect agreement
with the scaling law obtained from simulation.

\begin{table*}[ht]
\begin{center}
\begin{tabular}{c|cc|cc|cc|cc}
\hline
& \multicolumn{4}{c|}{Rings} & \multicolumn{4}{c}{Linear} \\
\hline
& \multicolumn{2}{c|}{Rouse Theory} & \multicolumn{2}{c|}{Simulation} & \multicolumn{2}{c|}{Rouse Theory} & \multicolumn{2}{c}{Simulation} \\
\hline
$N$ & $\tau_{D}/10^5$ & $g_1(\tau_{D})$ & $\tau_{D}/10^5$ & $g_1(\tau_{D})$ & $\tau_{D}/10^5$ & $g_1(\tau_{D})$ & $\tau_{D}/10^5$ & $g_1(\tau_{D})$ \\
\hline
 100 & 0.16 & 62.1  & 0.07     & 47  & 0.32 & 94.7   & 0.48  & 78  \\
 200 & 0.65 & 125.1 & 0.39     & 91  & 1.29 & 178.3  & 6.77  & 182 \\
 400 & 2.59 & 249.8 & 2.38     & 164 & 5.12 & 355.0  & 59.7  & 283 \\
 800 & 10.4 & 500.5 & 20.0     & 292 & 20.7 & 714.1  & 559.4 & --  \\
1600 & 41.4 & 998.6 & $>105.5$ & -- & 82.9 & 1429.2 & --    & --  \\
\hline
\end{tabular}
\end{center}
\caption{
Comparison of the diffusion times and mean-square displacement of the monomers between Rouse theory \cite{Doi86,Tsolou10} and the simulation results. The diffusion times are computed using $\tau_D=\langle R_g^2 \rangle/6D$. For the Rouse model $D=kT/N\zeta$ for both architectures while $\langle R_g^2 \rangle$ is $Nl_kl/6$ for the linear and $Nl_kl/12$ for the rings. To compute $\tau_D$ from the simulation data, $D$ is obtained from Table \ref{table3} and $\langle R_g^2 \rangle$ from Table II of the previous paper\cite{halverson_part1}.
The MD model is mapped to the Rouse model with $l_kl=2.71~\sigma^2$ and $\zeta=43~\tau^{-1}$.
 For the linear chain Rouse model $g_1(\tau_D)$ was computed using the middle of the chain (i.e., $n=N/2$). $\tau_D$ and $g(\tau_D)$ are given in units of $\tau$ and $\sigma^2$, respectively.
}
\label{table4}
\end{table*}

The peculiar conformational properties of the rings are expected to
lead to new dynamic properties. The ring melt simulations give a
number of unexpected and not fully understood results. For linear
polymers the overall stress relaxation and the crossover towards
free diffusion of the beads is linked to a motion distance of the
order of the radius of gyration, even though the center-of-mass
motion, $g_3(t)$, follows a $t^1$ power law at much shorter
displacements in comparison to the rings. This is due to the fact that any relaxation
is connected to the polymer diffusing out of its original tube. Such
a tube concept certainly cannot be applied for rings, at
least not directly. Also the strong connection of diffusion distance to
stress relaxation does not apply any more here without significant
alterations. In short, the dynamics can be summarized by the
following observations:

\begin{itemize} \item sub-diffusive behavior of $g_3(t) \sim t^y$ up to distances
2--3 times $\left\langle R_g^2 \right\rangle$ with $y \approx 0.75$, and a diffusion time
$\tau_D \sim N^x$, $x\geq 3$ (for $N=1600$ we even find a slowing down $g_3(t) \sim t^{y}$, 
$y\simeq 0.65$, for larger times),

\item the ``$t^{1/4}$'' regime in
$g_1(t)$ also extending 2--3 times $\left\langle R_g^2 \right\rangle$,

\item diffusion coefficient $D(N) \sim N^{-2.3}$ depends on $N$ in a similar way as in entangled
linear polymers, however the prefactor is approximately 7 times larger for the rings with $N = 400$ and 800 than the corresponding linear polymers,

\item zero-shear viscosity scaling
$\eta_0 \sim N^{1.4 \pm 0.2}$, which is much more suggestive of
Rouse-like behavior ($\sim N$) than reptation ($\sim N^{3.4}$), is
in accord with an internal reorganization time ${\cal O}(N^{2.2 \pm 0.1})$.
\end{itemize}

The first three points seem to contradict the fourth which suggests
that the overall motion of the chains and the stress relaxation, at
least to within our accuracy, are somewhat decoupled. In the
following we propose a mechanism which can account for this. Also,
the further slowing down of $g_3(t)$ for $N=1600$ is required in
order to prevent an early, unphysical crossing of $g_1(t)$ and $g_3(t)$
(see below).

At short times ($t \lesssim \tau_e$) we expect the motion of the beads, $g_1(t)$,
to follow the standard Rouse $t^{1/2}$ behavior, and as shown
in Fig. \ref{g13_rings_linear} this is indeed found. This is in contrast to $g_3(t)$ which does
not follow the standard Rouse behavior as discussed above. Once the entanglement time is
reached, a slowing down at a motion amplitude of the order of the
tube diameter $d_T^2 \approx b^2 N_{e,\mathrm{linear}}$ is expected.
This is observed for both the linear and ring polymers. Then one
expects for linear polymers the Rouse motion along the tube, leading
to the well-known $t^{1/4}$ regime and then, after the Rouse time, the
diffusion out of the tube, resulting in the second $t^{1/2}$ regime,
until after the disentanglement time $\tau_d \sim N^{3.4}$ where the
free diffusion dominates. For rings we expect a similar initial
slowing down due to the motion along the short branches of the
lattice animal-like structure (shorter rings) or of the crumpled 
globule (large rings). It appears, however, that
this slowed down motion and the
diffusion along the chain contour itself, as suggested by the decay
of the autocorrelation function of the cross product of the spanning
vectors, cf. Fig. \ref{relaxation_time}, create a unique power law
very close to $t^{1/4}$ over a very long time scale and distances
exceeding the gyration radius of the rings. Then there is a crossover towards free
diffusion without an additional clearly-defined power law in between, which is in contrast
to the behavior of linear chains.

Although we do not have a consistent theoretical explanation for the
observed dynamics, we can formulate an argument relating some of our
observations to the others.  Specifically, the slowed down
sub-diffusive motion of the beads requires some reduction in the
effective exponent of the center-of-mass motion as well, otherwise
$g_1(t)$ and $g_3(t)$ would meet or even cross (meaning that the coil as a whole
would move much further than a monomer, which is obviously
impossible) at a time even well below the Rouse time.  Let us assume
$g_3(t) \sim t^{\alpha_3}$ for the ``chain diffusion'' at
intermediate times and $g_1(t) \sim t^{\alpha_1}$ for one monomer
diffusion.  We might write for a time dependent ``diffusion
constant'' $D(g_1(t)) t^{\alpha_3} = g_1(t)$.  Assuming that $D(g_1(t))
\sim g_1(t)^{-d_f/2}$, namely proportional to the number of beads carried
along at a distance $g$ within an object of fractal dimension
$d_f=3$, we arrive at the relation $\alpha_3 = (5/2)
\alpha_1$\footnote{For a Rouse chain this argument shows that $g_3(t) \sim t$ requires $g_1(t) \sim t^{1/2}$ for $g_3(t)<\left\langle R_e^2 \right\rangle$.}.  This relation appears to be consistent with our
simulation data.  Indeed, careful inspection of the data for
$g_1(t)$ indicates that the smallest slope is close to $0.28$,
systematically decaying with increasing chain length to slightly
larger than $1/4$.  The smallest slope in $g_3(t)$
for $N=1600$ is close to $0.7$, which is in good agreement with the
above relation between $\alpha_1$ and $\alpha_3$.

Also the amplitude of $g_3(t)$, where the
crossover towards free diffusion occurs needs some more attention.
For instance, as shown in Fig. \ref{g13_rings_linear}, for the $N=400$
ring case the transition to $g_3(t) \sim t^1$ at long times occurs
at roughly $7.1\times10^5~\tau$ where $g_3(t)=170~\sigma^2$ or $3.2\langle
R_g^2 \rangle$. For the linear case at the transition one finds
$g_3(t)=170~\sigma^2$ or $0.9\langle R_g^2 \rangle$. For the ring
systems with $N=400$ and 800 (and extrapolated for $N=1600$) one then finds $D \sim N^{-2.3}$. All
previous simulation papers have reported a weaker dependence with
the strongest by Hur et al.\cite{Yoon2010} being $D \sim N^{-1.9}$. This is most
probably due to the use of effectively shorter chains in the
previous studies.

Another surprising result regarding the diffusion of the rings is
that the stress relaxation follows an approximate $t^{-1/2}$ scaling
law, in fair agreement with the $t^{-2/5}$ prediction of Kapnistos
et al.\cite{Rubinstein_Nature_2008} In fact, the $N=800$ ring
system shows this behavior for at least four decades in time. While
it is tempting to appeal to the Rouse model since it predicts $G(t)
\sim t^{-1/2}$, this certainly cannot apply. This fast decay,
compared to what we observe from the mean-square displacements of
the beads, is in accord with the observation that
$C_{\boldsymbol{c}\boldsymbol{c}}(t)$ decays with
a relaxation time $\tau_{\mathrm{relax}} \sim N^{2.2 \pm 0.1}$. That means
stress can relax by internal reorganization without the need to
diffuse a significant distance away from its original neighbor
chains or to fully relax the overall shape of the chain. This however is only possible because the rings eventually
are compact objects.
As a consequence, the characteristic stress
relaxation time $\tau_{\mathrm{relax}} \sim N^{2.2}$ implies that the
very small
viscosity $\eta_0$ and the diffusion time, $\tau_D \sim N^x$, $x\geq3$, are not directly
related to each other.
This line of arguments
does not take into account the existence of, as we called them,
quasi-entanglements, which relax only very slowly and might give
rise to a plateau-like regime in the stress relaxation function. For
a possible plateau, however, we expect a very small value of the
order ${\cal O}(k_BT/N)$, simply because the number of neighbors that a
given ring can entangle with becomes independent of $N$ and settles
at a value below 20, a typical number for a liquid. Thus, if
existing, such a plateau will be very difficult to observe and beyond
the accuracy of our data.

An attempt to describe the amoeba-like motion of the
individual ring among other nonconcatenated rings was presented in a
recent article by Milner and Newhall \cite{milner10}.  These authors
argue that since the ring in the lattice of obstacles passes each
lattice gate twice in the opposite directions and, therefore,
represents a kind of tree, then each bond separates the whole
system in two halves, with the sizes $k$ and $N-k$, respectively.
The smaller of the two numbers Milner and Newhall call ``centrality'' and
provide convincing simulation data to the effect that effective
motion along the ``reaction coordinate'' of centrality is
sub-diffusive with $\delta k \sim t^{3/4}$.  Clearly, the diffusion
of centrality in this picture is also closely related to the stress
relaxation, which means that this picture is ideologically very
close to that proposed by Kapnistos et al
\cite{Rubinstein_Nature_2008}. Given the identical values of the
exponents in the diffusion of centrality case and in our $g_3(t)$
behavior for intermediate times and ring sizes, it is tempting to 
bridge the two statements. However, our conformations
somehow suggest a rather extended regime of beads to qualify for the
centrality argument, making the precise meaning of this concept
somewhat unclear here. Future work
will be required to understand whether they are really related
or there is just a coincidence of numerical values for the two
exponents.

\section{Conclusions}
\label{sec:Conclusions}
The structural and dynamic properties of ring polymer melts
have been investigated by molecular dynamics simulation using
a semiflexible bead-spring model in this and the previous paper.
For the longest chain lengths, the radius of gyration was found to
approach a scaling regime of $N^{1/3}$ in agreement with previous
simulation results and theoretical studies. While this suggests collapsed conformations,
the rings are, in fact, open or expanded and highly interpenetrated by their
few ${\cal O}(N^0)$ neighbors.
This is supported by a primitive path analysis which was also used
to show that the estimated quasi-entanglement number of the rings is larger than that
of the analogous linear chains. The rings being not as entangled
as the linear chains and more compact in size may explain why they relax stress faster and have a viscosity which scales
much more weakly than the $N^{3.4}$ dependence of linear chains.
Despite following different scaling laws with $N$ for the radius of gyration, stress relaxation
and viscosity, and undergoing different modes of motion while diffusing, for large $N$ the
two architectures approximately obey the same $D \sim N^{-2.3 \pm 0.1}$ power law. Because
of the compact nature of the ring conformations, resulting in a fairly deep correlation hole,
obtaining diffusion constants from such simulations might be subject to rather strong finite
size effects in the number of chains considered. Such polymeric systems have some resemblance to a soft
sphere liquid which is known to be subject to such effects. However, we are confident that our systems with $M=200$ ring polymers are sufficiently large and we emphasize that studies using significantly fewer chains should be avoided.

Though we have presented a rather comprehensive analysis many questions
remain open. Conformational properties need further attention, especially
for questions with respect to biological issues. This concerns, for example,
contact maps and the life time of the clusters
seen in such maps\cite{halverson_part1}.
Also, the precise determination of $\beta$ and $\gamma$ and the related scaling
prediction $\beta + \gamma=2$ need further attention.
The decoupling of translational motion and stress relaxation
for rings, the latter leading to the low viscosity, certainly is not
fully understood and needs further investigation.
The sub-diffusive regime in the MSD 
of the rings up to displacements that are significantly larger than
$\left\langle R_g^2 \right\rangle$ poses special challenges for a proper
theoretical description.
 To better compare our findings to
experiments, a systematic study of the influence of linear
``contaminants'' is needed since experiments never can exclude
them completely. Such a study is in preparation\cite{Halverson10} and will also
include an extensive analysis of the dynamic scattering functions
of both the rings and the linear polymers.
Recent advances in synthesis, purification
and characterization should allow a better comparison in the near
future. While the main thrust of the two papers was on the properties
of nonconcatenated ring polymer melts, for comparison we also have
presented extensive simulations of highly entangled melts of linear
polymers. While in general the reptation picture provides a good
qualitative and semi-quantitative theoretical framework there are still
open questions with respect to the slow crossover to the anticipated
asymptotic regime or possible deviations thereof.

\begin{acknowledgments}
The authors are grateful to T. Vilgis, D. Fritz and V. Harmandaris
for their comments on an early version of the manuscript. The ESPResSo
development team is acknowledged for optimizing the simulation
software on the IBM Blue Gene/P at the Rechenzentrum Garching in
M\"{u}nich, Germany. We thank Donghui Zhang for discussions and
references relating to experimental studies on cyclic polymers. This
project was in part funded by the Alexander von Humboldt Foundation
through a research grant awarded to AYG. AYG also acknowledges
the hospitality of the Aspen Center for
Physics where part of this work was done. WBL acknowledges
financial support from the Alexander von Humboldt Foundation 
and the Basic Science Research Program through the National Research Foundation
of Korea (NRF) funded by the Ministry of Education, Science and
Technology (2010-0007886). Additional
funding was provided by the Multiscale Materials Modeling (MMM) initiative of the Max Planck
Society. We thank the New Mexico Computing Application Center NMCAC
for a generous allocation of computer time. This work is supported by
the Laboratory Directed Research and Development program at Sandia
National Laboratories. Sandia is a multiprogram laboratory operated
by Sandia Corporation, a Lockheed Martin Company, for the United
States Department of Energy under Contract No. DE-AC04-94AL85000.
\end{acknowledgments}


\end{document}